%% file: main.tex
\documentclass{article}
\usepackage{spconf,amsmath,graphicx,booktabs,enumitem}

\usepackage{siunitx}
\usepackage{subfiles}

\usepackage{tikz}

\title{Ensemble prosody prediction for expressive speech synthesis}

\name{
    \parbox{\linewidth}{
        \centering Tian Huey Teh$^{\dagger \star}$ \qquad Vivian Hu$^{\dagger}$ \qquad Devang S Ram Mohan$^{\dagger}$ \qquad   Zack Hodari$^{\dagger}$
        \mbox{Christopher G. R. Wallis}$^{\dagger}$ \quad Tomás Gómez Ibarrondo$^{\dagger}$ \quad Alexandra Torresquintero$^{\dagger}$ \quad James Leoni$^{\dagger}$ \\
        \mbox{Mark Gales}$^{\dagger \ddagger}$ \qquad \mbox{Simon King} $^{\dagger \mathsection}$}
    }

\address{
    $^{\dagger}$ Papercup Technologies Ltd., United Kingdom \\
    $^{\ddagger}$ University of Cambridge, United Kingdom \qquad
    $^{\mathsection}$ University of Edinburgh, United Kingdom \\
    \fontsize{9}{9}\selectfont\ttfamily\upshape
	tehtianhuey@gmail.com \qquad vivian@papercup.com \\
}

\begin{document}

\maketitle

\begingroup\renewcommand\thefootnote{\textsuperscript{*}}
\footnotetext{Work done while at Papercup Technologies Ltd.}
\endgroup

\newcommand\copyrighttext{%
  \footnotesize \textcopyright 2023 IEEE. Personal use of this material is permitted.  Permission from IEEE must be obtained for all other uses, in any current or future media, including reprinting/republishing this material for advertising or promotional purposes, creating new collective works, for resale or redistribution to servers or lists, or reuse of any copyrighted component of this work in other works.%
}
\newcommand\cpnotice{%
    \begin{tikzpicture}[remember picture,overlay]
    \node[anchor=south,yshift=10pt] at (current page.south){\fbox{\parbox{\dimexpr\textwidth-\fboxsep-\fboxrule\relax}{\copyrighttext}}};
    \end{tikzpicture}%
}

\begin{abstract}
\cpnotice%
Generating expressive speech with rich and varied prosody continues to be a challenge for Text-to-Speech. Most efforts have focused on sophisticated neural architectures intended to better model the data distribution. Yet, in evaluations it is generally found that no single model is preferred for all input texts. This suggests an approach that has rarely been used before for Text-to-Speech: an ensemble of models.

We apply ensemble learning to prosody prediction. We construct simple ensembles of prosody predictors by varying either model architecture or model parameter values. 

To automatically select amongst the models in the ensemble when performing Text-to-Speech, we propose a novel, and computationally trivial, variance-based criterion. We demonstrate that even a small ensemble of prosody predictors yields useful diversity, which, combined with the proposed selection criterion, outperforms any individual model from the ensemble.

\end{abstract}


\vspace{-2ex}
\section{Introduction}
\label{sec:intro}

Text-to-Speech (TTS) systems have become capable of producing synthetic speech with near-human naturalness. The focus has thus shifted towards generating varied and expressive speech that goes beyond ``average prosody'' \cite{hodari2019ssw10}. Typically, a new challenger model is proposed that demonstrably outperforms an incumbent model, by leveraging linguistic and context information \cite{mltb2022}, using reference audio \cite{prosospeech2022}, or adopting new architectures and losses \cite{portaspeech2021}. 

Yet, any newly-proposed model is almost never preferred by listeners \textit{all} the time. The incumbent model is still preferred  \textit{some} of the time. This motivates the use of multiple models in a way that combines their strengths. 

This idea is well established and is called \textit{ensemble learning}. Ensemble learning algorithms vary in their approach to two sub-problems \cite{ensemblesurvey2012}: i) \textit{ensemble generation}, the method used to construct multiple base learners, and ii) \textit{ensemble integration}, the method used to integrate their predictions, either by combining them (e.g., averaging) or selecting amongst them. Ensemble learning has been widely applied to Automatic Speech Recognition: \textit{system combination} constructs multiple systems which make complementary errors, then integrates their predictions using voting \cite{ROVER97} or likelihood-based re-scoring \cite{confusionnetwork2000}. 

In this paper, we construct ensembles of prosody predictors in order to increase the prosodic variety and expressiveness of the synthesised speech. In theory, a single generative model should be able to model the data distribution and thus generate varied samples. However, this has proved very difficult to achieve in practice. Because of the multi-faceted and probabilistic nature of prosody, different models --- influenced by their own inductive biases and the randomness inherent in model initialisation and training --- end up learning different aspects of the distribution. An ensemble is a convenient, practical solution to these problems \cite{abe2022}.

For ensemble generation, we investigate both \textit{homogeneous ensembles} comprising learners with the same architecture that differ only in their parameter values, and \textit{heterogeneous ensembles}, comprising learners of varying architectures. To demonstrate the idea, we limit the size of the ensemble to the minimum of two base learners. For ensemble integration, we select between the learners' predictions, rather than combining them (e.g., by averaging). This is to avoid exacerbating the ``average prosody" effect \cite{hodari2019ssw10}.

To our knowledge, this is the first use of ensemble learning with a selection approach to enhance expressivity in speech synthesis. Our contributions are to demonstrate that:

\vspace{-0.75em}
\begin{itemize}
\setlength\itemsep{-0.1em}
\item[i)] a minimal ensemble of just two prosody predictors yields sufficient perceivable diversity.\footnote{Samples: https://research.papercup.com/samples/\\ensemble-prosody-prediction} 
\item[ii)] an automatic criterion is able to select from the ensemble such that it outperforms any individual model.
\item[iii)] our best criterion, based on $F_0$ variance, is able to close the gap with a human upper bound by 31\%.
\end{itemize}

\section{Related work}
\label{sec:lit}

\subsection{Generative ensembles in speech synthesis}
\label{ssec:ensemble}

Prior work on ensemble learning for TTS is sparse. \cite{g2pensemble19} perform knowledge distillation on a token-level ensemble to improve the accuracy of grapheme-to-phoneme (G2P) conversion. They use a heterogeneous ensemble consisting of transformer, Bi-LSTM and convolution sequence-to-sequence models, and average the predicted probability distributions. As the output of G2P is discrete, prediction averaging is both appropriate and straightforward to apply.

\cite{prosodyensemble09} is closest to our work. They use the weighted average prediction of an ensemble of shallow neural networks to model $F_0$ and duration contours. They construct their ensemble using data cross-validation, creating six different subsets of training data, training eight fully connected networks on each subset that differ in the number of layers, hidden dimensions, and inputs. Our work differs in its focus on expressivity rather than accuracy, use of deep neural networks in ensemble generation, and use of selection rather than combination in ensemble integration. 


\vspace{-1ex}
\subsection{Oversmoothing and average prosody}

The issue of ``average prosody" in neural speech synthesis is analogous to oversmoothing in HMM speech synthesis, where statistical processing often led to overly smoothed generated parameter trajectories, causing muffled speech. \cite{GV07} proposed applying a global variance (GV) penalty to the likelihood objective of the parameter generation process, where GV measures the variance of the spectral and $F_0$ parameters over an utterance. They found GV to be inversely correlated with the oversmoothing effect and the penalty improved naturalness in subjective evaluation.

Drawing upon this, we explore the suitability of spectral GV and $F_0$ variance as criteria to select from ensembles of prosody predictors. 

\vspace{-1ex}
\subsection{Intonation}

$F_0$ is perhaps the most salient acoustic feature for modelling varied and expressive speech. \cite{hodari2019ssw10, hodari2020sp, wang2020} use variational autoencoder-based approaches with continuous or discrete latent variables to model intonation ($F_0$ trajectories). In these works, $F_0$ had to be extracted from the waveform. However, there has been a resurgence of explicit acoustic feature modelling, where acoustic features such as $F_0$ are explicitly predicted within the model \cite{fastspeech2020, raitio2020, fastpitch, CtrlP} by an \textit{acoustic feature predictor} (AFP). During inference, these acoustic features are predicted during the forward pass, then passed to the decoder for mel-spectrogram generation. As models of this type do generate speech with acoustic features matching the values predicted inside the model \cite{raitio2020, CtrlP}, we propose a measure of $F_0$ variance that leverages these explicitly predicted $F_0$ values. 

\input{figures/ensemble}

\section{Ensemble setup}
\label{sec:method}

\subsection{Model ensembles}
\label{ssec:model}
We demonstrate our proposed approach on ensembles of AFPs. All AFPs take as input the encoder outputs of the acoustic model and predict three speaker-normalised acoustic correlates of prosody at the phone level: $F_0$, energy, and duration.

We adopt two different architectures for the AFP: 

i) A \textbf{recurrent} neural network described in \cite{CtrlP}. This network consists of 4 Bi-LSTM layers with dimensions = 64, 64, 32, 32, which map encoder outputs to a sequence of hidden states. This is passed through a 16-dimensional fully connected layer, a tanh activation function, and a 3-dimensional projection layer. 

ii) A \textbf{convolutional} neural network based on the Temporal Predictor of FastPitch \cite{fastpitch}. This network consists of 2 stacked blocks, each comprising a 1-D convolution layer with kernel size 3 and filter size 256, a ReLU non-linearity, layer normalisation and dropout (p=0.1). The output of this stack is then passed through a final projection layer into 3-dimensions.

\label{sec:AFP_ensemble}
For each AFP architecture, we trained 2 versions that differ only in their parameter values (by using different random seeds for training). From these 4 checkpoints, we formed 3 ensembles of 2 members each. Two ensembles are homogeneous, comprising both checkpoints for a single architecture (\textbf{RNN-2} and \textbf{CONV-2}). The third ensemble is heterogeneous and contains one randomly-selected checkpoint for each architecture (\textbf{RNN-CONV}). We hypothesise that heterogeneous ensembles may yield more diverse prosody predictions than homogeneous ensembles, and benefit more from ensemble selection. 

Figure \ref{fig:ensemble} illustrates the ensemble of AFPs within the acoustic model. The acoustic model is the multi-speaker variant of Tacotron 2 \cite{taco2} which, when combined with
the recurrent AFP, becomes Ctrl-P \cite{CtrlP}. During inference, the AFP outputs are concatenated to the encoder outputs, attended over and decoded to generate a mel-spectrogram, which is input to a separate WaveRNN vocoder \cite{wavernn}.

\subsection{Selection criteria}
\label{ssec:criteria}
All our candidate selection criteria measure the variance of model predictions for the utterance being synthesised. Each criterion selects the rendition with the highest variance:

i) \textbf{GV}: following \cite{GV07}, we extract the first 25 mel-cepstral coefficients (MFCCs) from the acoustic model output mel-spectrogram (using librosa) and compute the average parameter-wise variance across the utterance, then sum those values to obtain the GV of that utterance;

ii) \textbf{WAV-F0} is the variance of $F_0$, as estimated from the synthetic speech waveform using RAPT \cite{rapt} at a frame shift of \SI{10}{\milli \second}, across all voiced frames; 

iii) \textbf{AFP-F0} is the variance of the $F_0$ values, as predicted by the AFP, across all voiced phones.

AFP-F0, GV and WAV-F0 are computed on the outputs of the AFP, acoustic model and vocoder, respectively. AFP-F0 only requires a forward pass of the encoder and the AFP, at which point the selection criterion can choose amongst the renditions proposed by the ensemble members. GV additionally requires a forward pass of the decoder per potential rendition. WAV-F0 adds waveform generation using the vocoder, again for each potential rendition. By operating on the output that is predicted earliest in the network, AFP-F0 takes 10x less computation than GV, and over 1000x less than WAV-F0.

\section{Experimental setup}
\label{sec:setup}

\vspace{-1ex}
\subsection{Training}
\label{ssec:training}
All models were trained on a proprietary expressive Mexican-Spanish corpus comprising a total of 38 hours of speech from 32 speakers. Voice actors were instructed to speak in a range of styles like "happy", "angry", "story telling" \ldots, resulting in wide variation in F0, energy, and duration. We followed the two-step training routine in \cite{CtrlP} in which the acoustic model is trained for 200k iterations, then frozen. Then, each AFP (Section \ref{sec:AFP_ensemble}) is trained for 400k iterations to minimise the L2 loss against ground truth acoustic feature values extracted from force-aligned reference speech. We used the Adam optimiser and a batch size of 16 in all stages. All experiments use the same WaveRNN checkpoint, trained for 3M iterations on mel-spectrograms generated in teacher forcing mode with ground truth acoustic feature values from an acoustic model of the same architecture.

\vspace{-1.5ex}
\subsection{Test materials}
\label{ssec:listening}
We used a test set of 30 sentences randomly selected from a compilation of videos (our target use case) that had been transcribed, translated and checked by human experts for semantic and linguistic accuracy. These sentences do not appear in the training data.

Our proposed method is designed to work for highly-expressive speech, so we selected 1 male and 1 female speaker which have among the widest coverage of stylistic variation (38 different styles) in the training data. For each speaker, we generate one pair of renditions per test sentence for each of the 3 model ensembles in Section \ref{ssec:model} (60 sample pairs in total per ensemble), to use in the listening tests.  


\vspace{-1.5ex}
\subsection{Perceived ensemble diversity}
\label{ssec:divsetup}
Ensemble diversity is a necessary prerequisite for any possibility of improvement. We evaluated this with 3 expert listeners who work with synthetic speech regularly and have a background in translation or linguistics. For each ensemble, we presented all utterance pairs, and asked ``Do A and B sound different? If yes, describe how they are different.", with choices `Yes' and `No', and an optional free-text field. 

\vspace{-1.5ex}
\subsection{Listener preference}
\label{ssec:prefsetup}
From the above responses, we identified the ensemble rated as most diverse: this is our best candidate for an ensemble-based TTS system. We conducted a second listening test to measure preference for the individual models in the ensemble using an A/B. 30 Mexican-Spanish native speakers crowd-sourced on the Prolific platform were asked (in Spanish) ``Choose which one you prefer: A or B?”, and given choices `A’, `B’ and `Undecided’. 

\vspace{-1.5ex}
\subsection{Selection criterion accuracy}
\label{ssec:accuracydefn}
We can then analyse these preference ratings to measure the ability of a selection criterion to exploit the ensemble's diversity. For each of the 3 criteria, we calculate \textit{accuracy} by comparing every of the 900 (30 sentences $\times$ 30 listeners) individual listener preferences with the choice of rendition made by that criterion. If they match, that criterion is considered accurate. 292 (16.2\%) of the responses were labeled `Undecided' and excluded from the accuracy calculation.





We also create a human \textbf{ORACLE}, by choosing the rendition of each utterance that was most preferred by listeners (i.e., the most frequent label amongst the 30 listener ratings). If ‘Undecided’ was the most frequent choice, all ratings for that utterance are excluded. ORACLE uses the majority vote of the crowd as the selection criterion, and represents the highest performance achievable by any selection criteria or individual model in this listening test. If all listeners agreed all the time, the ORACLE would achieve an accuracy of 100\%.

\input{tables/diversity}

\section{Results}
\label{sec:results}

\vspace{-1ex}
\subsection{Perceived ensemble diversity}
\label{ssec:diversity}

Table \ref{tab:diversity} reports the average amount of diversity perceived by each expert listener. The perceived diversity varies substantially, but it consistently shows that the heterogeneous ensemble RNN-CONV makes the most diverse prosody predictions. 

Table \ref{tab:keywords_vertical} summarises the free-text comments describing the perceived differences between renditions within each ensemble, which we tagged with keywords.
Most of the ensemble diversity arises from expressiveness, intonation, and emphasis, which we expect our selection criteria to correlate with. `Clarity' and `Speed/pauses' were often mentioned in tandem, with a high speech rate degrading clarity. These aspects of prosody will not be well captured by the criteria we explore because they relate to duration; future criteria could account for this. So, we expect our current selection criteria to exploit some, but not all, of the perceived diversity in the ensembles. 

\input{tables/keywords_vertical}

\vspace{-1.5ex}
\subsection{Listener preference}
\label{ssec:preference}

Figure \ref{fig:rnn-conv_ab} shows the preference results for the individual AFPs (\textbf{RNN}, \textbf{CONV}) in our most diverse ensemble. As expected, neither model is preferred all the time. Both models perform at chance level: i.e., listeners prefer each about half of the time, suggesting room to combine their strengths. We take RNN (as it is marginally preferred) as the non-ensemble baseline and compare all ensemble systems to that.

\vspace{-1.5ex}
\subsection{Selection criterion accuracy}
\label{ssec:accuracy}

Accuracy, as defined in Section \ref{ssec:accuracydefn}, is reported in Figure \ref{fig:accuracy}. 
Average preference for the non-ensemble baseline, RNN, is illustrated as a vertical dotted line.
We conducted two-sided Fisher’s exact tests to determine which accuracy rates were significantly different ($p \le 0.05$) from RNN, applying the Holm-Bonferroni correction. 

\input{figures/preference}

The top three bars show the results for the automatic selection criteria GV, WAV-F0 and AFP-F0. AFP-F0 has a significantly higher accuracy than RNN: automatic selection from an ensemble of prosody predictors is able to outperform its individual members. WAV-F0, which measures a similar property as AFP-F0, was not statistically different from RNN (unadjusted p-value: 0.094).
Counter-intuitively, GV is significantly \textit{worse} than RNN: that is, choosing the ensemble member with the \textit{lowest} GV would be a better criterion. GV is clearly measuring something useful from the generated mel-spectrograms, but exactly what is left for future work.

The last bar provides the human ORACLE upper bound accuracy defined in Section \ref{ssec:accuracydefn}; the value is less than 100\% because listeners, of course, do not perfectly agree in their preferences. 
AFP-F0 closes 31\% of the available performance gap between RNN and ORACLE, demonstrating that an automatic criterion is able to take advantage of the diversity present in an ensemble.

\vspace{-2ex}
\input{figures/accuracy}

\vspace{-2ex}
\section{Conclusions and future work}
\label{sec:end}
We have demonstrated that an ensemble of prosody predictors yields diversity that can be exploited by an automatic selection criteria, and thus outperforms any individual model. We demonstrated this gain with the smallest possible ensemble size of 2, an easy ensemble generation approach (two different architectures for a single component of the acoustic model), and a simple and computationally efficient $F_0$ variance selection criterion. There is a lot of future potential for exploring larger, more diverse ensembles, and more sophisticated selection criterion, especially regarding duration.

\bibliographystyle{IEEEbib}
\bibliography{refs}

\end{document}

%% file: figures/ensemble.tex
\begin{figure}[t]
    \centering
    \includegraphics[width=\columnwidth]{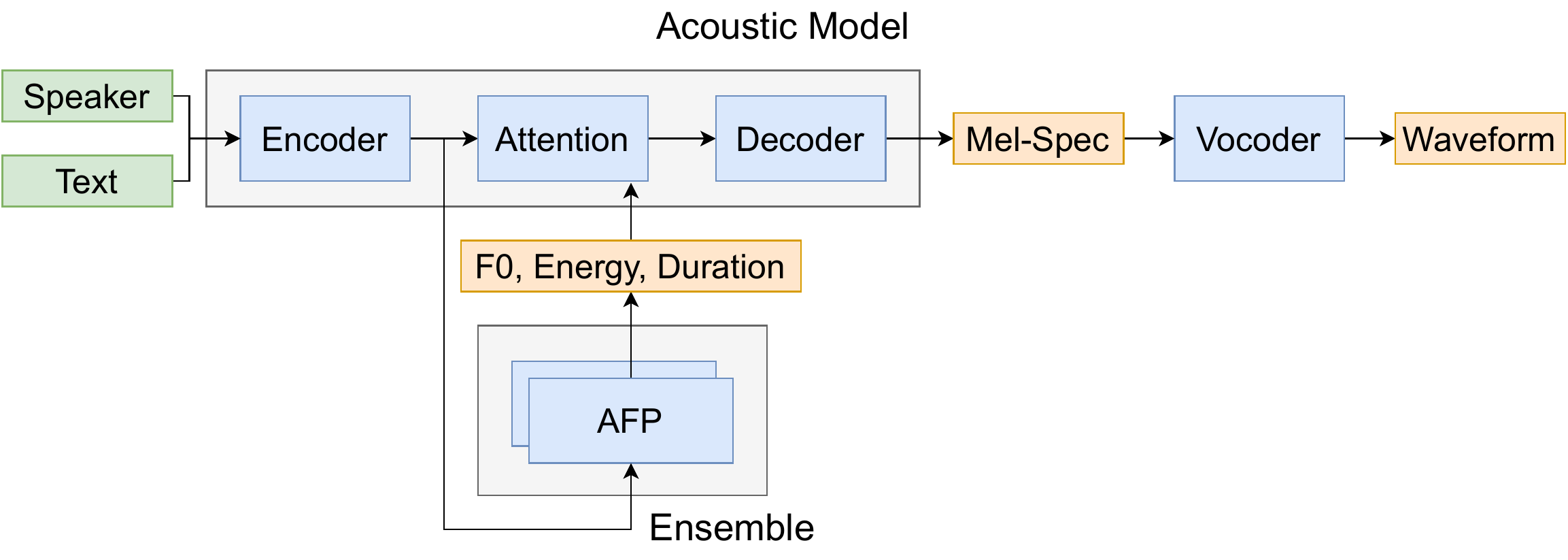}
    \caption{We consider ensembles of Acoustic Feature Predictors (AFPs), holding constant the acoustic model and vocoder.}
    \label{fig:ensemble}
\end{figure}

%% file: tables/diversity.tex
\begin{table}[t]
    \renewcommand{\arraystretch}{1.15}%
    \centering
    \resizebox{220pt}{!}{%
        \begin{tabular}{lccc}
            \toprule
            \multicolumn{1}{l}{\textbf{Ensemble}} & \multicolumn{1}{l}{\textbf{Listener 1}} & \multicolumn{1}{l}{\textbf{Listener 2}} & \multicolumn{1}{l}{\textbf{Listener 3}} \\
            \midrule
            \textbf{CONV-2} & 22\% & 40\% & 8\% \\
            \textbf{RNN-2} & 33\% & 57\% & 25\% \\
            \textbf{RNN-CONV} & 58\% & 70\% & 35\% \\
            \bottomrule
        \end{tabular}%
    }
    \caption{The percentage of utterance pairs rated as different is shown for each ensemble. All three expert listeners rated the heterogeneous ensemble (RNN-CONV) as most diverse. }
    \label{tab:diversity}
\end{table}


%% file: tables/keywords_vertical.tex
\begin{table}[b]
    \renewcommand{\arraystretch}{1.1}%
    \centering    
    \resizebox{220pt}{!}{%
        \begin{tabular}{lccc}
            \toprule
            \multicolumn{1}{l}{\textbf{Keyword}} & 
            \multicolumn{1}{l}{\textbf{CONV-2}} & 
            \multicolumn{1}{l}{\textbf{RNN-2}} & 
            \multicolumn{1}{l}{\textbf{RNN-CONV}} \\
            \midrule
            \textit{Expressiveness} & 7 & 11 & 27 \\
            \textit{Intonation} & 5 & 20 & 24 \\
            \textit{Emphasis} & 14 & 17 & 20 \\
            \textit{Clarity} & 3 & 9 & 16 \\
            \textit{Speed/pauses} & 9 & 14 & 25 \\
            \textit{Naturalness} & 3 & 9 & 13 \\
            \textit{Other} & 3 & 4 & 3 \\
            \bottomrule
        \end{tabular}%
    }
    \caption{Differences between utterance pairs frequently relate to expressiveness, intonation and emphasis. The table shows the count of keywords tagged in expert listeners' descriptive comments of the perceived differences in each ensemble.}
    \label{tab:keywords_vertical}
\end{table}


%% file: figures/preference.tex
\begin{figure}[t]
    \centering
    \includegraphics[width=225pt]{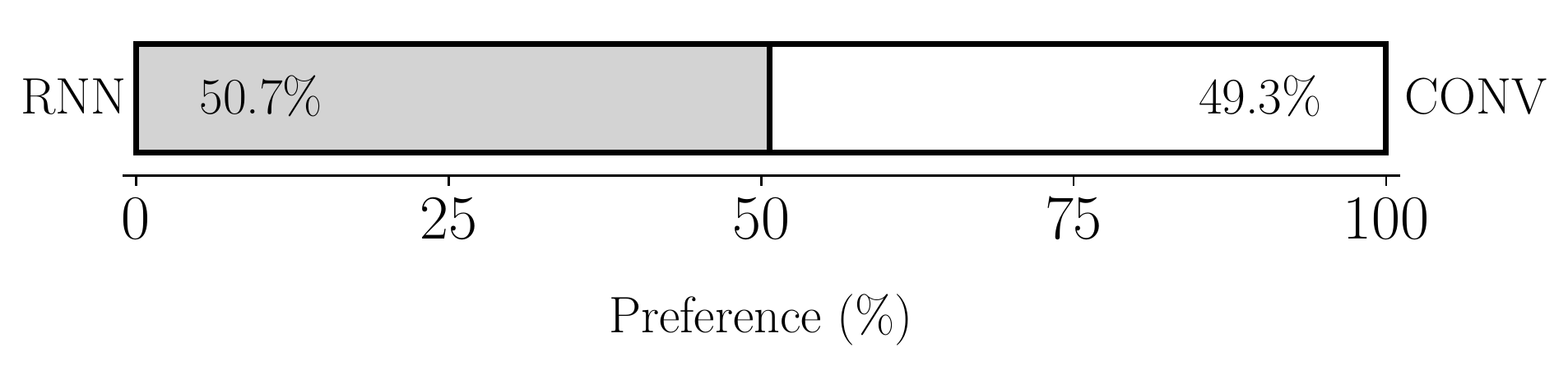}
    \vspace{-4pt}
    \caption{Both individual AFPs in our most diverse ensemble perform at chance level (two-sided binomial test, $p \le 0.05$)}
    \label{fig:rnn-conv_ab}
    \vspace{-4pt}
\end{figure}

%% file: figures/accuracy.tex
\begin{figure}[hb]
    \centering
    \includegraphics[width=220pt]{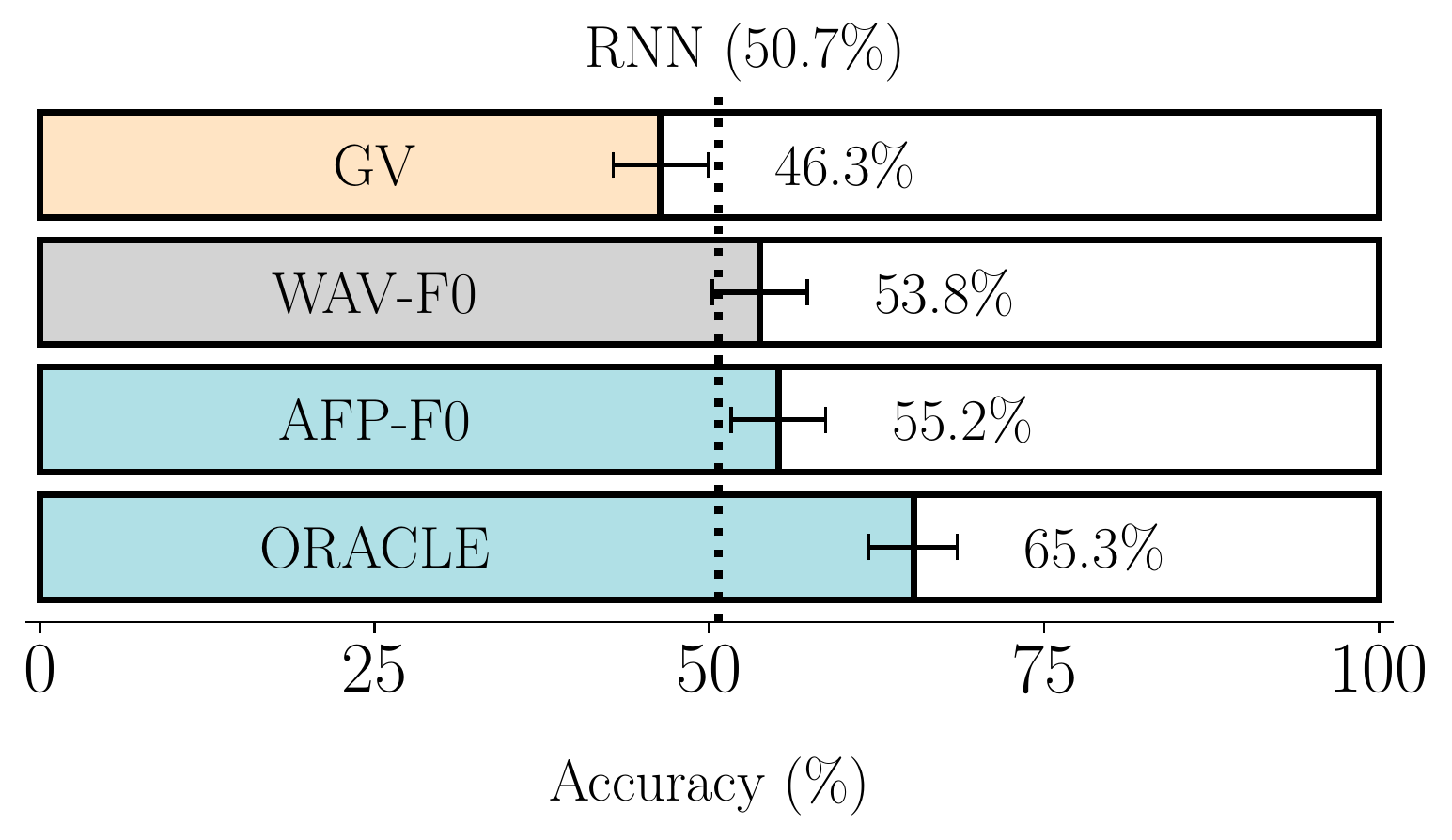}
    \vspace{-4pt}
    \caption{Our proposed criterion (AFP-F0) predicts listener preference more accurately than the single model RNN baseline (dotted line represents preference from Fig. \ref{fig:rnn-conv_ab}). Accuracy rates are shown with 95\% confidence intervals; blue and orange mark statistical significance.}
    \label{fig:accuracy}
    \vspace{-4pt}
\end{figure}



